\documentclass[article]{IEEEtran}
\usepackage{cite}
\usepackage{amsmath,amssymb,amsfonts}
\usepackage{algorithmic}
\usepackage{graphicx,color}
\usepackage{textcomp}
\usepackage{xcolor}
\usepackage{hyperref}
\hypersetup{hidelinks=true}
\usepackage{amsmath,amsfonts}
\usepackage{array}
\usepackage{amsmath,amssymb,amsfonts}
\usepackage{tikz}
\usepackage{pgfplots}
\usepackage{textcomp}
\usepackage{float}
\usepackage[ruled,linesnumbered]{algorithm2e}
\usepackage{subcaption}
\usepackage{comment}
\usepackage{color}
\usepackage{graphicx}
\usepackage{textcomp}
\usepackage{lettrine}
\usepackage{xcolor,flushend}
\usepackage{comment}

\newcommand{\floor}[1]{\left\lfloor #1 \right\rfloor}
\newcommand{\ceil}[1]{\left\lceil #1 \right\rceil}
\newtheorem{remark}{Remark}

\begin{document}

\title{Multi-Branch Attention Convolutional Neural Network for Online RIS Configuration with Discrete Responses: A Neuroevolution Approach\\

}
\author{
    George Stamatelis,~\IEEEmembership{Student~Member,~IEEE}, Kyriakos Stylianopoulos,~\IEEEmembership{Graduate~Student~Member,~IEEE}, \\and George C. Alexandropoulos,~\IEEEmembership{Senior~Member,~IEEE} \\

    \thanks{This work has been supported by the Smart Networks and Services Joint Undertaking (SNS JU) projects TERRAMETA and 6G-DISAC under the European Union’s Horizon Europe research and innovation programme under Grant Agreement No 101097101 and No 101139130, respectively. TERRAMETA also includes top-up funding by UK Research and Innovation (UKRI) under the UK government’s Horizon Europe funding guarantee.}
    \thanks{The authors are with the Department of Informatics and Telecommunications, National and Kapodistrian University of Athens, Panepistimiopolis Ilissia, 15784 Athens, Greece. G. C. Alexandropoulos is also with the Department of Electrical and Computer Engineering, University of Illinois Chicago, IL 60601, USA (e-mails: \{georgestamat, kstylianop, alexandg\}@di.uoa.gr).} 
}



\maketitle

\begin{abstract}
In this paper, we consider the problem of jointly controlling the configuration of a Reconfigurable Intelligent Surface (RIS) with unit elements of discrete responses and a codebook-based transmit precoder in RIS-empowered Multiple-Input Single-Output (MISO) communication systems. The adjustable elements of the RIS and the precoding vector need to be jointly modified in real time to account for rapid changes in the wireless channels, making the application of complicated discrete optimization algorithms impractical. We present a novel Multi-Branch Attention Convolutional Neural Network (MBACNN) architecture  for this design objective which is optimized using NeuroEvolution (NE), leveraging its capability to effectively tackle the non-differentiable problem arising from the discrete phase states of the RIS elements. The channel matrices of all involved links are first passed to separate self-attention layers to obtain initial embeddings, which are then concatenated and passed to a convolutional network for spatial feature extraction, before being fed to a per-element multi-layered perceptron for the final RIS phase configuration calculation. Our MBACNN architecture is then extended to multi-RIS-empowered MISO communication systems, and a novel NE-based optimization approach for the online distributed configuration of multiple RISs is presented. The superiority of the proposed single-RIS approach over both learning-based and classical discrete optimization benchmarks is showcased via extensive numerical evaluations over both stochastic and geometrical channel models. It is also demonstrated that the proposed distributed multi-RIS approach outperforms both distributed controllers with feedforward neural networks and fully centralized ones.
\end{abstract}


\section{Introduction}
The upcoming sixth-Generation (6G) of wireless networks is envisioned to support numerous new services with unprecedented user requirements, while achieving remarkable energy efficiency and immense reduction in network reconfiguration cost~\cite{nokia_whitepaper,6Gdisac}. Reconfigurable Intelligent Surfaces (RISs), and the resulting smart wireless environments, constitute one of the core candidate technologies for this vision~\cite{Huang_Reconfigurable_2019,RIS_arch2,towardsIRS} and are thus receiving significant attention from academia, industry, and standardization bodies~\cite{GR_001,GR_002}. An RIS comprises large numbers of metamaterial elements exerting electronically tunable responses to impinging electromagnetic waves, which can be programmed to dynamically control over the air the way information-bearing signals propagate~\cite{RISoverview2023_all}. Leveraging channel state observations~\cite{Swindlehurst_CE} and pertinent algorithms~\cite{RIS_arch1} in combination with minimal deployment and operation requirements~\cite{RIS_challenges_all}, RISs have the potential to transform a wide variety of wireless communication applications~\cite{risVehicular,RIS_counteracting_2023,RIS_loc,RIS_smart_city}.

To fully take advantage of the potential of RISs for wireless operations, precise real-time configuration of the internal states of their unit elements is necessary~\cite{rise6g_del26}. In practice~\cite{GR_002}, most algorithmic approaches demand the RIS to compute appropriate phase profile configurations and switch its elements' responses within the channel coherence time, which implies very stringent requirements in computational delays. Additionally, to facilitate the manufacturing process and reduce the associated costs, current RIS devices are being designed to admit only quantized phase profiles~\cite{RIS_challenges_all}, of a quantization of even a single bit~\cite{RIS_arch2,RIS_THz_terrameta}. Effectively, dynamically selecting RIS phase configurations reduces to discrete optimization problems that are typically associated with exponential computational complexities and sub-optimal solutions~\cite{RIS_arch1}. Despite the great attention RISs have recently attracted, the latter two challenges remain still largely unresolved, especially considering the large number of RIS elements, the dynamic nature of the wireless environments typically considered, as well as the computational requirements of the employed optimization algorithms.

\subsection{Background and Related Works}
To tackle the problem arising from the RIS discrete phase responses, one of the main options involves solving the continuous-phase problem and then discretizing the resulting configurations to the admissible phase states. In this respect, closed-form phase configurations guaranteeing at least a fraction of the achievable performance may be obtained~\cite{PZZ22, RIS_1bit_cf}, however, such approaches are not easily extensible to complex RIS systems, as envisioned under current 6G discussions (e.g.,~\cite{6Gdisac,RIS_THz_terrameta}). Different discrete optimization algorithms based on branch-and-bound methods~\cite{PZZ22, WZ20} or black-box genetic optimization schemes~\cite{PLP21, YCY23} have also been proposed, however, their computational complexity is prohibitively expensive for them to constitute realistic RIS phase configuration approaches in real-time (i.e., online) scenarios.

Acknowledging the latter challenge, the RIS research community has recently, yet extensively, studied the application of deep learning approaches as RIS configuration orchestrators~\cite{Huang_Indoor,Samarakoon_learning,RISnet,Jagyasi22_Unsupervised_RIS}. Despite the unavoidable overheads of data collection processes and immense computational cost required for training deep learning methods, they come with the advantage of extremely fast computation capabilities during the deployment phase, assuming though pertinent parallel hardware is available on the edge computational node that executes them. Deep learning techniques for RIS phase configuration fall under the supervised~\cite{Huang_Indoor,Samarakoon_learning,RISnet} or the unsupervised learning paradigm~\cite{Jagyasi22_Unsupervised_RIS}. However, such methods suppose a clear distinction between their training and deployment phases. As a result, the deployed Neural Networks (NNs) are not necessarily able to alter their decisions to account for environmental (i.e., channel) conditions not seen during training, which may lead to degrading performance as the distribution of the environmental stimuli deviates over time.

More practical for the problem of online RIS phase configuration control is the family of Deep Reinforcement Learning (DRL) algorithms. Approaches of this nature are specifically designed for perpetual training during deployment and are well-equipped to handle time-evolving systems. Due to the impressive success of DRL algorithms in games and robotic simulations, e.g., \cite{mnih2015humanlevel,alphaGo}, DRL methods for RIS phase configuration control have been extensively investigated in~\cite{Stylianop1Bit,DRL_RIS_secure,RIS_MISO_exploiting}, as well as in~\cite{AlexandroPervasive} and the references therein, establishing themselves as strong candidates for fast online RIS control solutions under diverse problems and systems. Notwithstanding their associated benefits, the implementation of DRL approaches comes with its own set of challenges.
At first, as discussed in~\cite{AlexandroPervasive,stylianop-MAB-ICC}, the vast majority of RIS phase profile control problems only marginally align with the paradigm of Markov Decision Processes (MDPs), which is the cornerstone of reinforcement learning. In fact, the MDP formulations implicitly assume that current RIS control actions affect future states of the wireless channels, however, this does not apply in practical formulations.
MDP algorithms are nevertheless equipped with mechanisms, like advantage estimation\cite{schulman2018highdimensionalcontinuouscontrolusing} and eligibility traces\cite{Sutton1998}, to tackle such time evolution dependencies. Therefore, applying ``off-the-shelf'' MDP solutions
to RIS control problems, due to ease of implementation, incurs unnecessary complications that increase the computational and training overheads as well as stability and generalization. The multi-armed bandit framework has been investigated as a candidate methodology in~\cite{AlexandroPervasive, stylianop-MAB-ICC, other-RIS-MAB} to remedy for this, however, state-of-the-art schemes are not easily extensible to large RISs with complicated action spaces. 

NeuroEvolution (NE) approaches constitute simpler alternatives than reinforcement learning for training NNs in dynamic settings. Using NE to solve MDPs or similar decision making tasks is a relatively old idea~\cite{moriarty:mlj96, introtoEA}, however, it has been recently shown that it can compete or outperform state-of-the-art DRL algorithms~\cite{such2018deep,salimans2017evolution}. Based on evolutionary and genetic algorithms~\cite{introtoEA} to train policy NNs, a major benefit of NE approaches relies on the fact that they can be parallelized during training over multiple computing nodes with each node only needing to share few scalar values~\cite{salimans2017evolution}. More importantly, since NE methods are gradient-free, they largely avoid instability issues that often plague typical deep learning and DRL optimization schemes associated with backpropagation. This feature is extremely advantageous for RIS  control in the light of large numbers of elements with quantized phase responses, which constitutes a problem that is not naturally compatible with differentiation-based algorithms. However, to our best of knowledge, whether and how NE can be effectively utilized for online RIS control tasks remains still unexplored and is one of the focal points of this paper. 

Notwithstanding the available training framework of deep learning or DRL-like approaches, the second important challenge with efficient online RIS phase configuration control is the architectural design of NNs. Typical DRL methods employ NNs that receive all Channel State Information (CSI) as input observations in the format of concatenated vectors or image-like tensors, thus, relying on immense numbers of NN parameters to extract useful patterns from such large unstructured data. There has been recent research effort in designing tailored NN architectures that take advantage of internal properties of the CSI matrices for more efficient feature extraction, predominately capitalizing on the impressive capabilities of the ``attention'' mechanism~\cite{AttentionAllYouNeed} that has enabled highly performing ``transformer'' architectures for sequence modeling~\cite{nueralMachineTranslation}. The ChanFormer NN was designed in~\cite{chanformer} to handle the decomposition of the channel matrices into its multi-path components. A similar architecture for spectrum forecasting and CSI estimation was used in~\cite{transformerSpectrum} and~\cite{attention-CE}, respectively, while~\cite{StylianopTRacking} proposed an attention architecture for user tracking with an RIS-based Receiver (RX).

\subsection{Contributions}
In this paper, we focus on the online orchestration of RIS-empowered Multiple-Input Single-Output (MISO) communication systems considering a more realistic, challenging discrete version than the one studied by prior works, such as \cite{DRL_RIS_secure,RIS_MISO_exploiting,Stylianop1Bit, AlexandroPervasive}. We develop a novel NN tailored for such systems which we efficiently optimize with an NE approach. Our contributions are summarized as follows:
\begin{enumerate}
     \item We present a novel Multi-Branch Attention Convolutional NN (MBACNN) architecture for the online joint configuration of the RIS discrete phase responses and the Transmitter (TX) precoding vector. The different sub-modules of the proposed NN are designed to capture different types of patterns appearing in the channel matrices and impose correlations in the elements of the selected RIS configurations.
     \item To effectively deal with the considered design problem which is non-differentiable, we introduce a simple NE framework to train the proposed MBACNN architecture. 
		  \item We extend the MBACNN architecture to multi-RIS-empowered MISO communication systems and present an NE-based optimization approach for the online distributed configuration of the phase profiles of the RISs.  
			\item Our extensive numerical investigations over both stochastic and geometrical channels demonstrate 
that our MBCANN architecture is superior to existing DRL schemes, lightweight discrete optimization benchmarks, as well as simple-evolved Feed Forward (FF) NNs with vastly larger numbers of trainable parameters. In addition, it is showcased that our online distributed optimization of multiple RISs outperforms large centralized optimizers. It is also interestingly shown that the proposed NE-based training procedure is relatively robust to various changes in its hyper-parameters.
 \end{enumerate}

The remainder of the paper is organized as follows. Section~II includes our system model and problem formulation, and Section~III presents our MBCANN architecture and its evolution for the online design of RIS-empowered MISO communication systems. Section~IV extends this architecture and NE-based joint optimization to  multiple RISs, while Section~V presents our extensive performance evaluation results. The paper is concluded in Section~VI.

\subsection{Notation}
Bold letter notation refers to vectors, e.g., $\mathbf{x}$, and bold capitals to matrices, e.g., $\mathbf{X}$, while calligraphic capital letters denote sets, e.g., $\mathcal{X}$. $\mathbb{R}$ and $\mathbb{C}$ are the real and complex number sets. $\mathbf{0}_{n}$ and $\mathbf{I}_{n}$ ($n\geq2$) represent the $n$-element all-zeros vector and the $n\times n$ identity matrix, respectively. The $(i,j)$-th element, the $i$-th row, and the $j$-th column of $\mathbf{X}$ are respectively denoted by $[\mathbf{X}]_{i,j}$, $[\mathbf{X}]_{i,:}$, and $[\mathbf{X}]_{:,j}$, while $\mathbf{X}^{\rm T}$ and $\mathbf{X}^{\rm H}$ represent $\mathbf{X}$'s transpose and conjugate transpose, respectively. The expectation operation is denoted as $E[\cdot]$, whereas $\mathfrak{Re}\{\cdot\}$ and $\mathfrak{Im}\{\cdot\}$ return respectively the real and the imaginary part of a complex quantity. $x\sim\mathcal{N}(\mu,\sigma^2)$ and $x\sim\mathcal{CN}(\mu,\sigma^2)$ indicate that the random variable $x$ is respectively normal and complex normal with mean $\mu$ and variance $\sigma^2$. Finally, $\jmath\triangleq\sqrt{-1}$ is the imaginary unit.

\section{System Model and Problem Formulation}\label{sec:system}
In this section, we introduce the considered RIS-empowered MISO communication system and present our online design objective for its reconfigurable parameters.

\subsection{System Model with A Single RIS}
Consider an $N_{\rm TX}$-antenna TX wishing to establish a data communication link with a single-antenna RX.
To strengthen this link, an RIS consisting of $N_{\rm RIS}$ phase-tunable unit elements is deployed in between the end nodes \cite{Moustakas_Cap2023}, so that its area of influence can alter their end-to-end Signal-to-Noise Ratio (SNR)~\cite{GR_001,RIS_challenges_all}. Following the majority of the available RIS proofs of concept \cite{RIS_arch2,GR_001}, all metametamterials are assumed to realize two-state tunable phase shifts on the impinging signals (i.e., $1$-bit phase resolution).
We model the considered RIS-empowered MISO communication as a sequence of channel coherence blocks, indexed by $t$, so that the Channel Frequency Response (CFR), modeled as a random variable, remains quasi-static within each block and changes independently and identically distributed (i.i.d.) from one block to the next one.
In particular, we model the phase response of each $i$-th ($i=1,2,\ldots,N_{\rm RIS}$) RIS unit element at each time instant $t$ via the parameter $\phi_i(t) \in [-1, 1]$. Accounting for $1$-bit quantization, we choose $\theta_1=-1$ and $\theta_2=1$ as our admissible $\phi_i(t)$ values $\forall t$; the extension of our framework to more discrete phase states will be discussed in the sequel. We define $\boldsymbol{\phi}(t) \triangleq [\phi_1(t)\, \phi_2(t)\,\cdots\, \phi_{N_{\rm RIS}}(t)]^{\rm T}$ as the phase configuration vector including all $\phi_i(t)$'s of the RIS.
Therefore, the RIS phase configuration matrix can be defined as $\mathbf{\Phi}(t)\triangleq{\rm diag}[\boldsymbol{\varphi}(t)]$ with ${\boldsymbol{\varphi}}(t)\triangleq[\exp(\jmath\pi \phi_1(t))\,\exp(\jmath\pi \phi_2(t))\,\cdots\,\exp(\jmath\pi \phi_{N_{\rm RIS}}(t))]$.
In baseband representation, the complex-valued received signal at the RX at each $t$-th time instant can be expressed as~\cite{Huang_Reconfigurable_2019}:
\begin{equation}
\label{eq:setUP}
    y(t) \triangleq \left(\mathbf{h}^{\rm H}(t)+\mathbf{h}_2^{\rm H}(t) \mathbf{\Phi}(t) \mathbf{H}_1^{\rm H}(t)\right) \mathbf{v}(t) {x}(t) + \tilde{n}(t),
\end{equation}
where $\mathbf{h}(t) \in \mathbb{C}^{N_{\rm TX} \times 1}$ includes the channel gain coefficients between the single RX antenna and the TX antennas, $\mathbf{H}_1(t) \in \mathbb{C}^{N_{\rm TX} \times N_{\rm RIS}}$ represents the channel matrix between the RIS and the TX, and $\mathbf{h}_2(t) \in \mathbb{C}^{N_{\rm RIS}\times 1}$ is the channel vector between the RX and the RIS. The TX precoding vector $\mathbf{v}(t) \in \mathbb{C}^{N_{\rm TX} \times 1}$ is assumed selected from a discrete codebook $\mathcal{V}\triangleq\{ [\mathbf{V}]_{:,i}\}_{i=1}^{N_{\rm TX}}$, whose elements correspond to the columns of the $N_{\rm TX} \times N_{\rm TX}$ Discrete Fourier Transform (DFT) matrix $\mathbf{V}$. Note that this set of precoders is a simplified version of the 3GPP 5GNR Type I codebook~\cite{3GPP_TS_38.214_codebooks}.
In addition, at each time instant $t$, ${x}(t)$ denotes the complex-valued transmitted data symbol with power $E[|x(t)|^2]=P$, and $\tilde{n}(t)\sim\mathcal{CN}(0,\sigma^2)$ is the Additive White Gaussian Noise (AWGN). We herein make the common assumption that AWGN's variance $\sigma^2$ can be reliably estimated; it is thus assumed to be perfectly known.

We assume coherent communications for the considered RIS-empowered MISO system, which indicates that, for each time instant $t$, all the involved channels $\mathbf{h}(t)$, $\mathbf{H}_1(t)$, and $\mathbf{h}_2(t)$, need to be accurately estimated to then design $\mathbf{v}(t)$ and $\mathbf{\Phi}(t)$. It is noted that both the channel estimation as well as the joint TX precoding and RIS phase configuration design are known to be quite challenging tasks~\cite{RIS_arch1}, resulting in control signaling~\cite{RIS_CC} and computation~\cite{wu2021intelligent} overheads that also depend on the operation capabilities of the RIS and its control architecture~\cite{RISoverview2023_all,rise6g_del26}, especially when realistic considerations are made for the TX precoders and the RIS phase profiles~\cite{Stylianop1Bit,AlexandroPervasive,RIS_1bit_cf}. Hence, when all the latter channels appearing in the received signal model in~\eqref{eq:setUP} are perfectly known, the instantaneous end-to-end SNR of the considered system can be easily obtained as follows:
\begin{equation}
\label{eq:SNR-DEF}
\gamma(t) \triangleq \frac{P}{\sigma^2}\left|(\mathbf{h}^{\rm H}(t)+\mathbf{h}_2^{\rm H}(t) \mathbf{\Phi}(t) \mathbf{H}_1^{\rm H}(t)) \mathbf{v}(t)\right|^2.
\end{equation}

\subsection{Design Problem Definition} \label{sec:problem-definition}
In this paper, 
we aim to find the combination of the RIS phase configuration vector and the discrete TX precoding vector that maximizes the instantaneous SNR, i.e., to solve:
\begin{equation*}
\begin{split}
  \mathcal{OP}_1: &\max_{\boldsymbol{\phi}(t),\mathbf{v}(t)} \quad\gamma(t)
	\\& \hspace{0.6cm}\textrm{s.t.}~~\quad\phi_i(t) \in \{\theta_1,\theta_2\} \,\,\forall i, \hspace{2.8cm}({\rm C1})
	\\& \hspace{1.24cm}  \quad\mathbf{v}(t) \in \mathcal{V}.  \hspace{4.283cm}({\rm C2}) 
\end{split}
\end{equation*}
This problem formulation is traditionally solved via iterative discrete optimization approaches due to the quantization limitation of constraint $({\rm C1})$ and the non-continuous decision variable of constraint $({\rm C2})$. However, the discrete nature of both constraints makes the problem NP-hard, which poses the following interconnected challenges: \textit{i}) The usually employed algorithms require exponential computational complexity that scales with the number $N_{\rm RIS}$ of RIS elements. 
\textit{ii}) The instantaneous nature of the control problem implies that the execution of the optimization algorithm should be completed well within the channel coherence time, which is expected to be in the order of few milliseconds.
This attribute of real-time decision making is crucial in the foreseen RIS-enabled smart wireless environments, since computation-induced latency may not only lead to outdated CSI at the time of configuration, but also to reduced spectral efficiency due to the unavoidable idle time depriving data transmission.

We proceed by defining the mapping function $g(\cdot)$ that is intended to satisfy $\mathcal{OP}_1$'s objective. In particular, it maps the instantaneous channel matrices to an RIS phase configuration and a TX precoding vector, i.e.: 
\begin{equation}\label{eq:mapping-function}
    \{\boldsymbol{\phi}(t),\mathbf{v}(t)\} = g\left(\mathbf{h}(t), \mathbf{H}_1(t),\mathbf{h}_2(t)\right).
\end{equation}
As previously mentioned, to address the computational limitations associated with iteratively solving $\mathcal{OP}_1$ at every time instance $t$, we require the computational requirements of the mapping function to be small. Typically, a linear computational complexity, i.e., $O(N_{\rm RIS})$, along with parallelizable operations and supporting hardware are reasonable assumptions to achieve computational latency in the order of one millisecond, or below, for moderate RIS sizes. By using~\eqref{eq:mapping-function}, it is convenient to re-express the SNR in~\eqref{eq:SNR-DEF} as:
\begin{align}\label{eq:SNR-re-def}
    \gamma\left(\boldsymbol{\phi}(t), \mathbf{v}(t)\right) \triangleq \gamma\left(g\left(\mathbf{h}(t), \mathbf{H}_1(t),\mathbf{h}_2(t)\right)\right),
\end{align}
to emphasize the role of the mapping/decision function $g(\cdot)$. To this end, $\mathcal{OP}_1$ can be equivalently written as follows:
\begin{equation*}
\begin{split}
  \mathcal{OP}_2: &\max_{g(\cdot) \in \mathcal{G}} \quad\gamma\left(g\left(\mathbf{h}(t), \mathbf{H}_1(t),\mathbf{h}_2(t)\right)\right)
	\\& \hspace{0.33cm}\textrm{s.t.}~\hspace{0.3cm}({\rm C1})\,\,{\rm and}\,\,({\rm C2}), 
\end{split}
\end{equation*}
where $\mathcal{G}$ represents the set of all applicable mappings. Note that the output $\{ \boldsymbol{\phi}(t),\mathbf{v}(t)\}$ of $g(\cdot)$ is used for the evaluation of the end-to-end SNR and the optimization problem is now over functions rather than the TX precoding and RIS phase configuration vectors.
In principle, $\mathcal{G}$ may contain any function capable of mapping to the feasible sets of the RIS phase configurations and TX precoding vectors.
In practice, families of heavily parameterized statistical models, such as NNs, can be utilized, leveraging their high expressivity capabilities, parallelizable training gains, fast inference times, as well as the option for
designing their internal modules at high level as ``architectural priors'' to take advantage of known patterns or structures in the input/output variables.

Solving $\mathcal{OP}_2$ at every time instant $t$ involves searching over over-parameterized function domains. This is still rather intractable despite our intention for computationally tractable $g(\cdot)$ mappings. To alleviate from the computational cost of per-time-instant optimization, we propose to relax $\mathcal{OP}_2$ to its online version, as follows. Consider a finite time horizon $T$, so that $t=1,2,\ldots,T$.
At every time instant $t$, we invoke \eqref{eq:mapping-function} to design $\{ \boldsymbol{\phi}(t),\mathbf{v}(t)\}$ upon observing the current CSI (see, e.g., \cite{RIS_arch1,Swindlehurst_CE,RIS_Vahid_low,HRIS_CE_all} and references therein, for CSI estimation schemes for RIS-empowered systems). Our new objective is to maximize the expected sum-SNR over the time horizon, which is expressed mathematically as follows:
\begin{equation*}
\begin{split}
  \mathcal{OP}_3: &\max_{g \in \mathcal{G}} \quad\frac{1}{T}\mathbb{E}\left[ \sum_{t=1}^{T}\gamma\left(g\left(\mathbf{h}(t), \mathbf{H}_1(t),\mathbf{h}_2(t)\right)\right) \right]
	\\& \hspace{0.33cm}\textrm{s.t.}~\hspace{0.3cm}({\rm C1})\,\,{\rm and}\,\,({\rm C2}).
\end{split}
\end{equation*}
In this formulation, the AWGN and the channel matrices are treated as Random Values (RVs) and the expectation is taken over their joint probability density function over the $T$ time instants. Since the joint distributions involved are intractable, the expectation can be treated via optimizing $g(\cdot)$ over a sequence of Monte Carlo episodes of independent RV samples. 

It can be easily noticed that, at an arbitrary time instant $t$, the values of the RVs at all time steps $t'$ with $t < t' \leq T$ cannot be known, therefore, it is not possible to solve $\mathcal{OP}_3$ for the whole duration of the episode. Instead, one may only start with an arbitrary selection of $g(\cdot)$, proceed to use it to compute the SNR through \eqref{eq:SNR-re-def} at any time instant $t$, and, finally, update the mapping function accordingly toward maximizing $\gamma(t)$. Treating the objective as a summation of per-time-instant objectives has the advantage of $g(\cdot)$ being optimized online, i.e., over time, as more data become available. Leveraging the requirements for a computationally fast $g(\cdot)$ , the total computational cost of the optimization is amortized per time instant $t$, leading to minute computational overheads per $t$.

The previously described time-horizon-based problem formulation and solution approach encompass the notion of learning, where, given effective training procedures and mapping functions, and after a sufficient amount of time instants have elapsed, the design $g(\cdot)$ function outputs effective RIS phase configurations and TX precoding vectors with small latency at each channel coherence time. Obviously, for $T \to \infty$, $\mathcal{OP}_1$ and $\mathcal{OP}_3$ become equivalent, since the fraction of time spent for training becomes negligible. Notwithstanding, even in finite-horizon problems, the online maximization formulation presents a trade-off between training overhead and decision latency. The goal of online learning procedures, such as DRL, NE, and unsupervised learning is to find near-optimal decision functions as quickly as possible.

\section{MBACNN Architecture and Deployment}\label{sec:arch}
In this section, we first introduce our MBACNN architecture tasked to learn the mapping between wireless channels and the configuration of the RIS phase profile together with the TX precoding vector. We then present an NE scheme for the training of the proposed learning approach, which is followed by a discussion on the MBACNN deployment for the considered RIS-empowered MISO communication system.    

\subsection{Proposed MBACNN Architecture}\label{sec:MBACNN_arch}
\begin{figure*}
    \centering
    \begin{subfigure}[]{0.4\textwidth}
    \centering
        \scalebox{0.55}{\includegraphics{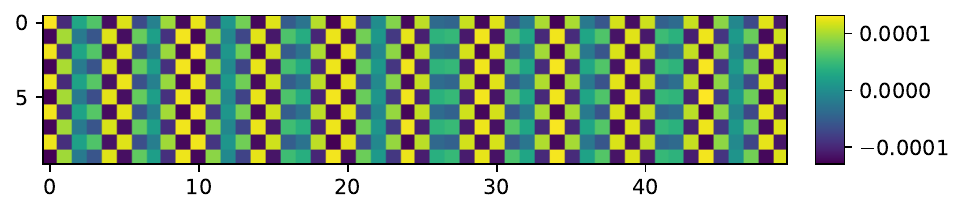}}
    \caption{Real part of a $\mathbf{H}_1(t)$ realization.}
    \end{subfigure}
    \begin{subfigure}[]{0.4\textwidth}
    \centering
        \scalebox{0.55}{\includegraphics{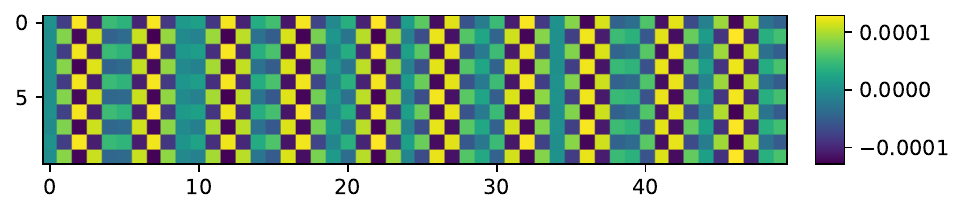}}
    \caption{Imaginary part of a $\mathbf{H}_1(t)$ realization.}
    \end{subfigure}
    \caption{The real and imaginary parts of a $10\times50$ channel matrix $\mathbf{H}_1(t)$ constituting a realization of the Ricean distribution \cite[eq. (5)]{AlexandroPervasive} with $7$ dB $\kappa$-factor. It can be observed that adjacent matrix elements have similar values (spatial correlation), a fact that motivates the investigation of attention mechanisms as a means to extract important channel features.}
    \label{fig:riceanSeqDemo}
\end{figure*}

Inspired by the potential of RISs, and their subsequent optimization, in low angular spread scenarios~\cite{Moustakas_Cap2023,AlexandroPervasive}, we illustrate in Fig.~\ref{fig:riceanSeqDemo} the real and imaginary parts of a realization of a $10\times50$ channel matrix $\mathbf{H}_1(t)$ (i.e., between a $10$-antenna TX and an RIS with $50$ elements) drawn from the Ricean channel model in \cite[eq. (5)]{AlexandroPervasive} with a moderate $\kappa$-factor at $7$ dB. It can be observed that adjacent columns of the matrix have similar values due to the induced spatial correlation~\cite{Nakagami_corr,Bjornson2020Rayleigh} arising from the steering vector expressions. In fact, each column is strongly dependent on neighboring ones, implying that the columns of the channel matrix form patterns. This fact motivates the consideration of sequence modeling tools~\cite{attentionSurvey,AttentionAllYouNeed,nueralMachineTranslation}  to extract important information regarding the wireless channels.

\begin{figure*}
    \centering
    \scalebox{0.7}{\includegraphics{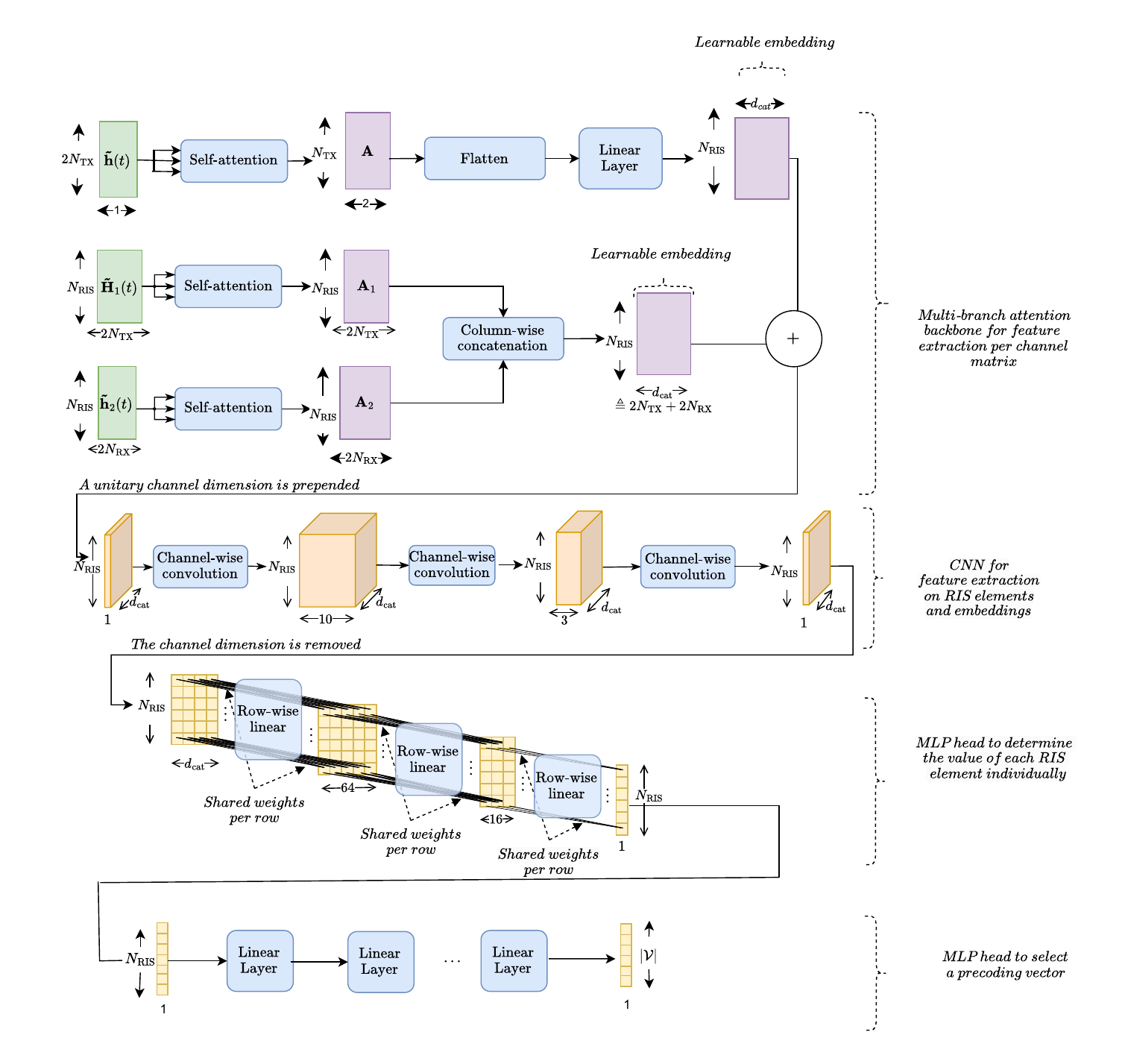}}
    \caption{The proposed MBACNN architecture comprising a multi-branch attention module, followed by a Convolutional Neural Network (CNN) module, a Multi-Layered Perceptron (MLP) module for the RIS phase configuration selection, and an additional MLP module for the TX precoding vector selection.}
    \label{fig:dACNN}
\end{figure*}
The proposed MBACNN architecture is illustrated in Fig.~\ref{fig:dACNN} and is comprised of four main modules, each intended to capture different types of correlations and patterns in the structure of the input channel matrices and the output active (at the TX) and  passive/reflective (at the RIS) beamforming configurations. In the following, we will discuss each of these modules in detail. At first, to handle complex operations, before being passed to the proposed NN, the wireless channels are transformed as follows:
\begin{align}
 \tilde{\mathbf{h}}(t)& \triangleq [\mathfrak{Re}\{\mathbf{h}^{\rm T}(t)\}\,\mathfrak{Im}\{\mathbf{h}^{\rm T}(t)\}]^{\rm T} \in \mathbb{R}^{2N_{\rm TX}\times 1},\\
    \tilde{\mathbf{H}}_1(t)&\triangleq\begin{bmatrix}
        \mathfrak{Re}\{\mathbf{H}_1(t)\}\\
        \mathfrak{Im}\{\mathbf{H}_1(t)\}
    \end{bmatrix} \in \mathbb{R}^{2N_{\rm TX} \times N_{\rm RIS}},\\
    \tilde{\mathbf{h}}_2(t)&\triangleq [\mathfrak{Re}\{\mathbf{h}^{\rm T}_2(t)\}\,\mathfrak{Im}\{\mathbf{h}^{\rm T}_2(t)\}]^{\rm T} \in \mathbb{R}^{2N_{\rm RIS}\times 1}.
\end{align}

\subsubsection{Attention Module}
The first module in the proposed MBACNN architecture illustrated in Fig.~\ref{fig:dACNN}, that receives as inputs the channels $\tilde{\mathbf{h}}(t)$, $\tilde{\mathbf{H}}_1(t)$, and $\tilde{\mathbf{h}}_2(t)$ at each time instant $t$ and is tasked for feature extraction, is based on self-attention~\cite{AttentionAllYouNeed,attentionSurvey}. By viewing each channel matrix as a sequence of (row) vectors, this module seeks patterns among the channel coefficients related to the RIS unit elements. The sequence comprises of RIS elements, and each element is represented as a token vector containing the coefficients at the TX and RX antennas. Self-attention architectures are efficient in extracting sequence-related information~\cite{nueralMachineTranslation}, in effect, unveiling the channel conditions that induced the correlations between the channel coefficients at adjacent RIS elements.  

As shown in the first module in Fig.~\ref{fig:dACNN}, each of the channels $\tilde{\mathbf{h}}(t)$, $\tilde{\mathbf{H}}_1(t)$, and $\tilde{\mathbf{h}}_2(t)$ is passed on its separate self-attention layer in order to capture important features regarding the correlations of the channel matrix elements. Their learnable parameters are intended to weigh the correlations so that the NN identifies which correlations are useful for the active/passive beamforming problem at hand. Consider for example the channel matrix $\tilde{\mathbf{H}}_1(t)$, for which we remove the time variable notation for simplicity in the following expressions. The first term in implementing neural self-attention for $\tilde{\mathbf{H}}_1$ is to compute the attention scores as follows:
\begin{equation}\label{eq:attention}
    \mathbf{S}_1 \triangleq {\rm softmax}\Big( \frac{\mathbf{Q}_1 \mathbf{K}_1^{\rm T}}{\sqrt{2N_{\rm TX}}}  \Big) \in \mathbb{R}^{N_{\rm RIS} \times N_{\rm RIS}},
\end{equation}
where ${\rm softmax}(\cdot)$ is taken over the complete array to convert all values in the range $[0,1]$, and matrices $\mathbf{Q}_1$ and $\mathbf{K}_1$ are:
\begin{align}\label{eq:attention-Q-K}
    \mathbf{Q}_1 &\triangleq \tilde{\mathbf{H}}_1 \mathbf{W}^q_1, \quad 
    \mathbf{K}_1 \triangleq \tilde{\mathbf{H}}_1 \mathbf{W}^k_1 \in \mathbb{R}^{2N_{\rm TX} \times 2N_{\rm TX}},
\end{align}
where $\mathbf{W}^q_1$ and $\mathbf{W}^k_1$ are real-valued learnable weight matrices of dimension $N_{\rm RIS} \times 2 N_{\rm TX}$. Note that the matrix multiplication inside \eqref{eq:attention} may be interpreted as row-wise dot products that measure the row similarity for all pairs of rows in $\tilde{\mathbf{H}}_1$. The weight matrices perform task-specific linear transformations on these matrix row pairs. The final, attended output of each attention layer is given as follows:
\begin{equation}\label{eq:attention-output}
    \mathbf{A}_1 \triangleq \mathbf{S}_1 \mathbf{P}_1 \in \mathbb{R}^{ N_{\rm RIS} \times 2 N_{\rm TX}},
\end{equation}
where we have used the matrix definition:
\begin{equation}\label{eq:attention-V}
    \mathbf{P}_1 \triangleq \tilde{\mathbf{H}}_1 \mathbf{W}^v_1 \in \mathbb{R}^{N_{\rm RIS} \times 2 N_{\rm TX}},
\end{equation}
with $\mathbf{W}^v_1 \in \mathbb{R}^{N_{\rm RIS}\times 2 N_{\rm TX}}$ being trainable and having the role to further measure the relevance between the attention scores and the task-specific learnable information embedded in its weights. Following similar reasoning, the attended outputs $\mathbf{A}_2 \in \mathbb{R}^{N_{\rm RIS} \times 2}$ and $\mathbf{A} \in \mathbb{R}^{N_{\rm TX} \times 2}$ resulting after respectively passing $\tilde{\mathbf{h}}_1$ and $\tilde{\mathbf{h}}$ through appropriate self-attention branches are constructed. The matrices $\mathbf{A}_1$ and $\mathbf{A}_2$ are concatenated along their column dimension in the matrix $\mathbf{A}_{\rm C} \triangleq {\rm colcat}(\mathbf{A}_1, \mathbf{A}_2) \in \mathbb{R}^{N_{\rm RIS} \times d_{\rm cat}}$ with $d_{\rm cat} \triangleq 2N_{\rm TX} + 2$.  

In order to be able to merge all the matrices before passing them to the next module in the proposed MBACNN architecture, an additional FF layer with the appropriate number of hidden units is utilized to transform $\mathbf{A}$ (first flattened to an $2N_{TX} \times 1$ vector) into the $\mathbf{A}_0 \in \mathbb{R}^{d_{\rm cat} N_{\rm RIS} \times 1}$ matrix. This matrix is then reshaped to the dimension $N_{\rm RIS} \times d_{\rm cat}$ and the final attended matrix is obtained as follows:
\begin{equation}\label{eq:A_T}
\mathbf{A}_{\rm T} \triangleq \mathbf{A}_{\rm C}+\mathbf{A}_0,    
\end{equation}
where layer normalization needs to be performed at each of the two terms of the summation.

It is finally noted that, in many RIS-focused investigations (e.g.,~\cite{Stylianop1Bit,RIS_challenges_all,RIS_MISO_exploiting,RIS_CC,PartiallyConnectedRIS,Huang_Reconfigurable_2019}), the direct TX-RX link, via the channel $\mathbf{h}(t)$, is considered as blocked or highly attenuated. This case can be easily treated by our MBACNN architecture; the architecture in Fig.~\ref{fig:dACNN} simplifies by removing the attention head that receives as input the $\tilde{\mathbf{h}}(t)$ vector.

\subsubsection{CNN Module}
The next module for feature extraction is based on cascaded convolutional layers. To this end, the attended matrix $\mathbf{A}_{\rm T}$ formulated in \eqref{eq:A_T} is treated as an image-like tensor of dimension $1 \times N_{\rm RIS} \times d_{\rm cat}$, where a single image channel is implied during the first dimension. Three convolutional layers are applied to the input tensor, with the hyper-parameters configured so that the last two output dimensions of each layer remain unchanged. Instead, the channel dimension (i.e., number of kernels) differs among the layers and is used to allow the NN to store arbitrary information. The output of the final convolutional layer again consists of a single channel, which is then discarded to construct a matrix of the same dimension as $\mathbf{A}_{\rm T}$. 

The convolutional layers of this module are intended to extract spatial patterns in the ``image-like'' stacked attention matrix, as well as to decrease the dimensionality of the input. In fact, notwithstanding rich scattering conditions, channel matrices exhibit high spatial locality in terms of phase values and possibly amplitude, therefore, the use of convolutional kernels is expected to be highly effective in detecting them.

\subsubsection{Feed Forward Layers for the RIS Phase Configuration $\boldsymbol{\phi}(t)$}
Having tasked the first two modules to detect patterns among the channel coefficients at the RIS elements as well as at the TX antennas, the third module of the proposed MBACNN architecture in Fig.~\ref{fig:dACNN} is a Multi-Layered Perceptron (MLP) designed to map the output of the CNN module to discrete RIS phase configurations. The same stack of $\tanh(\cdot)$ activated linear layers is applied to each of the $d_{cat}$ rows outputted by the CNN module. The shape of the layers is selected such that a single number is outputted, and by applying it to each row, an $N_{\rm RIS}$-element RIS phase configuration vector $\bar{\boldsymbol{\phi}}(t)$ is obtained. From the definition of the $\tanh(\cdot)$ activation function, the elements of $\bar{\boldsymbol\phi}(t)$ lie in the interval $(-1,1)$. Therefore, the entire phase configuration vector is passed through a ${\rm sign}(\cdot)$ function to get binary values (specifically, $1$ or $-1$), finally, leading to the desired $\boldsymbol{\phi}(t)$. \begin{remark}[Extension to Arbitrary Numbers of $\phi_i(t)$ Values]
The output dimension of the final layer of the third module of the proposed
MBACNN architecture can be increased to the number of the available discrete phase states; this extension treats cases with more than two $\phi_i(t)$ values $\forall i$~\cite{RIS_challenges_all,RIS_THz_terrameta}. For such cases, the ${\rm softmax}(\cdot)$ activation function needs to be used in the place of the $\tanh(\cdot)$ and ${\rm sign}(\cdot)$ combination.
\end{remark}
 
\subsubsection{Feed Forward Layers for the TX Precoder $\mathbf{v}(t)$}
The next task of the proposed MBACNN architecture is to select the most appropriate precoding vector from the available ones in the discrete codebook $\mathcal{V}$ solving $\mathcal{OP}_3$. To do so, we deploy one additional stack of linear layers activated by a Rectified Linear Unit (ReLU) that maps $\mathbf{v}(t)$ to an $|\mathcal{V}|\times 1$ output (with $|\cdot|$ providing the set cardinality). This output is then passed through a ${\rm softmax}(\cdot)$ activation function to generate a discrete distribution over all possible TX precoding vectors. We then sample a precoding vector index from that distribution.


\subsection{Proposed NE-Based Optimization}\label{sec:NE}
As mentioned earlier, one of the key challenges of the design problem considered in this paper (i.e., $\mathcal{OP}_3$) is the discrete nature of the optimization variables, as imposed by the constraints $({\rm C1})$ and $({\rm C2})$, which makes differentiation-based algorithms, such as backpropagation\cite{backprog}, inapplicable.
To deal with this difficulty, we adopt the NE optimization paradigm~\cite{benchmaringBasics} to train the proposed NN that is tasked to find the optimal $\boldsymbol{\phi}(t)$ and $\mathbf{v}(t)$ solving $\mathcal{OP}_3$.
Specifically, we deploy the COperative SYnamptic NEuroevolution (CoSyNe) scheme that was initially proposed in~\cite[Algorithm 2]{cosyne}, which is briefly described below.

Let us first define the abstract parameter vector $\mathbf{w} \in \mathbb{R}^{M\times1}$ including all $M$ trainable parameters of the NN by vectorizing and concatenating in an arbitrary order the respective parameter matrices of the individual layers presented in Section~\ref{sec:arch}.
Subsequently, we re-express the mapping function in \eqref{eq:mapping-function} as $g_{\mathbf{w}}\left(\mathbf{h}(t), \mathbf{H}_1(t),\mathbf{h}_2(t) \right)$ to make the NN's dependence on its weight values explicit. The CoSyNE scheme keeps a set of $L_{\rm pop}$ different instances of $\mathbf{w}$ as a population of candidate solutions that maximize the fitness functions. By representing the population with $\boldsymbol{W} \in \mathbb{R}^{L_{\rm pop} \times M}$, each $\ell$-th row $[\boldsymbol{W}]_{\ell,:}$ with $\ell = 1,2,\ldots, L_{\rm pop}$ constitutes an {\em individual} parameter vector consisted of $M$ {\em chromosomes}. 
Moreover, CoSyNE defines each $m$-th column $[\boldsymbol{W}]_{:,m}$ with $m=1,2,\ldots, M$ as a {\em subpopulation} of chromosomes. Note that the notion of a subpopulation represents a collection of candidate values for each chromosome of an individual parameter vector.

The CoSyNE-based algorithm solving $\mathcal{OP}_3$ first initializes $\boldsymbol{W}$ randomly, and then repeats the following steps at every generation until convergence:
\begin{enumerate}
    \item For each $\ell$-th individual with $\ell=1,2,\ldots,L_{\rm pop}$, obtain the parameter vector $[\boldsymbol{W}]_{\ell,:}$ and compute its fitness score, $f_\ell$, 
    using the NN $g_{[\boldsymbol{W}]_{\ell,:}}(\cdot)$.
    \item Sort the rows of $\boldsymbol{W}$ according to their fitness scores $f_\ell$'s in descending order.
    \item Use the top $\floor{L_{\rm pop}/4}$ rows as parents and construct $\ceil{3L_{\rm pop}/4}$ offspring, denoted by $\mathbf{w}^{\rm o}_{\ell}$ for $\ell=\ceil{3L_{\rm pop}/4},\ceil{3L_{\rm pop}/4}+1,\ldots, L_{\rm pop}$, through standard crossover and mutation mechanisms.
    \item Replace the last $\ceil{3L_{\rm pop}/4}$ rows of $\boldsymbol{W}$ as $[\boldsymbol{W}]_{\ell,:} \gets \mathbf{w}^{\rm o}_{\ell}$ for $\ell=\ceil{3L_{\rm pop}/4},\ceil{3L_{\rm pop}/4}+1,\ldots, L_{\rm pop}$.
    \item Assign to each chromosome $[\boldsymbol{W}]_{\ell,m}$ a probability of permutation as $p^{\rm perm}_{\ell,m} = 1- \sqrt[M]{f_\ell/f_{\rm max}}$ $\forall \ell,m$, where $f_{\rm max}$ is the highest fitness score among the individuals of the generation.
    \item With probability $p^{\rm perm}_{\ell,m}$ $\forall \ell,m$, mark each chromosome for permutation.
    \item For each $m$-th subpopulation with $m=1,2,\ldots,M$, randomly shuffle all chromosomes of the column $[\boldsymbol{W}]_{:,m}$ that have been marked for permutation.
\end{enumerate}
The main difference between CoSyNE and standard genetic optimization algorithms~\cite{introtoEA} is that the former further incorporates the concept of co-evolution, where, by randomly permuting less-fit chromosomes among candidate solutions, a more efficient search is conducted through the parameter space, as compared to solely mutating and generating offspring. We next describe the crossover procedure. Suppose two individuals $l_1$ and $l_2$ are marked as parents, and that their offspring's chromosomes will replace those of an arbitrary individual $l'$ in $\boldsymbol{W}$.
Each chromosome of $l'$ is then sampled according to the following distribution $\forall$$m=1,2,\ldots,M$:
\begin{equation}\label{eq:crossover}
    [\boldsymbol{W}]_{l',m} = \begin{cases}
        [\boldsymbol{W}]_{l_1,m} &, \text{probability } 0.5\\
        [\boldsymbol{W}]_{l_2,m} &, \text{probability } 0.5
    \end{cases}.
\end{equation}
The mutation mechanism involves stochastically updating each chromosome of a new offspring $l'$ (with $\eta \sim \mathcal{N}(0,\sigma_{\rm mut}^2)$):
\begin{align}\label{eq:mutation}
    [\boldsymbol{W}]_{l',m} &= \begin{cases}
        [\boldsymbol{W}]_{l',m} + \eta&, \text{probability } p_{\rm mut} \\
        [\boldsymbol{W}]_{l',m} &, \text{probability } 1 - p_{\rm mut} 
    \end{cases}.
\end{align}

Computing the fitness function for an individual NN is done over a sample of $T_E$ episodes, each of which has horizon $T$. For each time instant $t \leq T$, we sample (i.e., estimate) the channels $\mathbf{h}(t)$, $\mathbf{H}_1(t)$, and $\mathbf{h}_2(t)$, which are then provided to the NN in order to output $\mathbf{\Phi}(t)$ and $\mathbf{v}(t)$. The instantaneous SNR is calculated and stored.  For a candidate mapping $g_{\mathbf{w}}()$, the total fitness function is the average instantaneous SNR expressed as:
\begin{equation} \label{eq:fitness-function}
    f \triangleq \frac{1}{T_E}\sum_{t_e=1}^{T_E} \frac{1}{T} \sum_{t=1}^T \gamma( g_\mathbf{w}(\mathbf{h}(t),\mathbf{H}_1(t),\mathbf{h}_2(t)) ).
\end{equation}
It is noted that~\eqref{eq:fitness-function} is a sample-based approximation of $\mathcal{OP}_3$'s objective function, which motivates the methodology of episodic learning. A pseudocode describing the fitness function calculation, which is needed for computing each $f_\ell$ in the previously described Step $1$ of our CoSyNE-based algorithm, is available in Algorithm~\ref{alg:fitness}. 
\SetKwInput{KwData}{Input}
\begin{algorithm}[!t]
\caption{Fitness Function Calculation}\label{alg:fitness}
\KwData{Horizon $T > 0$, candidate MBACNN $g_{\mathbf{w}}(\cdot)$, noise level $\sigma^2$, and TX power level $P$.}
Set $f \gets 0$. \\
\For{$t_e=1,2,\ldots,T_E$}{
\For{$t=1,2,\ldots,T$}{
Obtain estimates for $\mathbf{h}(t)$, $\mathbf{H}_1(t)$, and $\mathbf{h}_2(t)$.\\
Obtain $\{ \mathbf{\Phi}(t),\mathbf{v}(t) \} = g_{\mathbf{w}}\left(\mathbf{{h}(t)},\mathbf{H}_1(t),\mathbf{h}_2(t)\right)$.\\ Set 
$f \gets f + \gamma\left(g_{\mathbf{w}}\left(\mathbf{h}(t), \mathbf{H}_1(t),\mathbf{h}_2(t)\right)\right)$.
}}
Set $f \gets \frac{f}{T_E T}$.\\
Return $f$.
\end{algorithm}

\section{Extension to Multi-RIS-Empowered Systems} \label{sec:distr}
In this section, we extend the MBACNN architecture presented in Section~\ref{sec:MBACNN_arch} to multi-RIS-empowered MISO communication systems, and present an NE-based joint optimization approach for the TX precoding design and the distributed configuration of the phase profiles of the multiple RISs.

\subsection{System Model with Multiple RISs}
Consider a smart wireless environment, within which the point-to-point MISO communication system is operating, comprising $K$ identical RISs, and let $\mathbf{H}_{1,k}(t)$ and $\mathbf{h}_{2,k}(t)$ indicate the channel gain matrices between the TX and each $k$-th RIS (with $k=1,2,\ldots,K$) and between each $k$-th RIS and the RX, respectively. 
The baseband received signal model in~\eqref{eq:setUP} now generalizes to the following expression:
\begin{equation}
\label{eq:received_signal_distributed}
y(t)= \mathbf{m}(t)\mathbf{v}(t) x(t) +\tilde{n}(t),
\end{equation}
where $\mathbf{m}(t)$ represents the multi-RIS-parameterized channel:
\begin{equation}
    \mathbf{m}(t)\triangleq\mathbf{h}^{\rm H}(t)+\sum_{k=1}^{K} 
    \mathbf{h}_{2,k}^{\rm H}(t)
    \mathbf{\Phi}_k(t) \mathbf{H}_{1,k}^{\rm H}(t)
\end{equation}
with $\mathbf{\Phi}_k(t)$ being the diagonal phase configuration matrix of each $k$-th RIS. For notation purposes, let $\boldsymbol{\phi}_k(t)$, with its $i$-th element denoted by $\phi_{i,k}(t)$, represent the phase configuration
vector for each $k$-th RIS, similar to the definition of $\boldsymbol{\phi}(t)$ in Section~\ref{sec:system}. Subsequently, the end-to-end SNR in~\eqref{eq:SNR-DEF} generalizes to $\gamma(t) = \frac{P}{\sigma^2}|\mathbf{m}(t)|^2$ and the mapping $g(\cdot)$ in~\eqref{eq:mapping-function} to:
\begin{equation}\label{eq:mapping-function_multi}
    \left\{\{\boldsymbol{\phi}_k(t)\}_{k=1}^K,\mathbf{v}(t)\right\} = g\left(\mathbf{h}(t), \{\mathbf{H}_{1,k}(t)\}_{k=1}^K,\{\mathbf{h}_{2,k}(t)\}_{k=1}^K\right).
\end{equation}
It is noted that, in the multi-RIS system model in~\eqref{eq:received_signal_distributed}, the channel paths involving two or more RIS-induced reflections have been ignored since the associated multiplicative pathloss results in negligible contribution to the arriving signal at the RX, and the precise modeling and estimation of the RIS-to-RIS links remain yet unsolved challenges~\cite{PhysFad}.


\subsection{Centralized vs. Distributed RIS Optimization}\label{cen_vs_dis}
One of the core design decisions one has to take for solving $\mathcal{OP}_3$'s extension to the multi-RIS case, i.e., the following design optimization problem for the system parameters:

\begin{equation*}
\begin{split}
  \mathcal{OP}_4: &\max_{g \in \mathcal{G}} \,\frac{1}{T}\mathbb{E}\!\left[ \sum_{t=1}^{T}\!\gamma\!\left(g\!\left(\mathbf{h}(t), \{\mathbf{H}_{1,k}(t)\}_{k=1}^K,\{\mathbf{h}_{2,k}(t)\}_{k=1}^K\right)\!\right)\!\right]
	\\& \hspace{0.06cm}\textrm{s.t.}~\hspace{0.3cm}\phi_{i,k}(t) \in \{\theta_1,\theta_2\} \,\,\forall i,k\,\,{\rm and}\,\,({\rm C2}),
\end{split}
\end{equation*}
is whether a centralized or a distributed approach will be used for the phase configuration optimization of the $K$ RISs. If a central processing unit (e.g., a Graphics Processing Unit (GPU)) for all RISs were to be used to solve $\mathcal{OP}_4$ (most presumably at the TX~\cite{RIS_challenges_all}), the instantaneous CSI for all the channels would need to be provided to it via dedicated control links of non-negligible capacity~\cite{RIS_CC}. Additionally, those control links should support bi-directional communications for the transmission of the $\mathcal{OP}_4$-optimizing phase configuration vectors $\boldsymbol{\phi}_k(t)$'s by the central processor to each $k$-th RIS controller. Evidently, this design would entail a large overhead in terms of control signaling and synchronization. Additionally, the dimensions of both the input and the output matrices in the proposed MBACNN architecture would exhibit a $K$-fold increase, 
so that the mapping function between such high-dimensional spaces might be too complicated to efficiently approximate and optimize.
Finally, even during inference time, the size of the NN may be too computationally demanding to support real-time control.

Suppose that a distinct processing unit is being used to decide the phase configuration vector $\boldsymbol{\phi}_k(t)$ for each $k$-th RIS (i.e., solve $\mathcal{OP}_4$ for each RIS separately). One example, though still far from the capabilities of the current RIS prototypes (see \cite{ETSI_RIS_mag} and references therein), could be that each $k$-th RIS is a hybrid active/passive metasurface~\cite{HRIS_Mag_all}, or even a reconfigurable intelligent computational surface (RICS)~\cite{RICS}, capable of also of channel estimation at its side~\cite{HRIS_CE_all} (thus, avoiding over-the-air CSI sharing as in the centralized approach) through a processing unit that can be computationally boosted to also decide $\boldsymbol{\phi}_k(t)$. In this distributed setup, policy approximation could be easier \cite{zhang2021multiagentreinforcementlearningselective}, since each distinct processing unit would be equipped with a lighter NN. To this end, each distributed processing unit could use an identical copy of the optimized NN to learn the mapping function for a single RIS. In the sequel, we present our extended MBACNN architecture for distributedly designing the phase configuration vectors of the multiple RISs, and discuss its advantages over the previously sketched centralized approach. 

\subsection{Proposed Distributed NE-Based Optimization}\label{sec:NE_multi}
The proposed framework for designing the phase configurations of the $K$ RISs in a distributed manner consists of a lightweight messaging protocol and a decentralized parameter sharing evolution approach based on the CoSyNE algorithm~\cite{cosyne}. In particular, we enrich the previously designed MBACNN architecture with only an additional MLP module to distributedly decide each $\boldsymbol{\phi}_k(t)$. For convenience, for the proposed NN represented by $g_\mathbf{w}(\cdot)$ tasked to solve $\mathcal{OP}_4$, we henceforth use notation $g_\mathbf{w}^{(1:4)}$ to represent its first four modules, as introduced in Section~\ref{sec:arch}, and notation $g_\mathbf{w}^{(5)}(\cdot)$ to indicate the newly added module.  

The proposed distributed setting for solving $\mathcal{OP}_4$ involves $K+1$ agents: \textit{i}) Each of the first $K$ agents is tasked with controlling each single $k$-th RIS, i.e., deciding each $\boldsymbol{\phi}_k(t)$. As previously mentioned, each agent could be implemented by each computationally boosted RIS controller (i.e., as in the extended hybrid RIS case or RICS) or by a device capable for this computation placed next to a conventional RIS controller. \textit{ii}) The final $(K+1)$-th agent, which is placed at the RX side, is responsible for deciding the TX precoding vector $\mathbf{v}(t)$, and consequently, feeding it back to the TX. More specifically, this agent receives lightweight messages from the $K$ RIS agents, which are passed to a separate NN as side information so that $\mathbf{v}(t)$ is finally determined. Concretely, the MBACNN parameters $g_\mathbf{w}^{(1:4)}(\cdot)$ of the first four modules at all $K$ RIS agents are shared with the $(K+1)$-th agent, which is tasked to map $\mathbf{h}(t)$, $\mathbf{H}_{1,k}(t)$, and $\mathbf{h}_{2,k}(t)$ to each $k$-th RIS phase configuration $\boldsymbol{\phi}_k(t)$, and decide a precoding vector index from $1$ to $|\mathcal{V}|$, which is henceforth denoted by $\bar{v}_k(t)$. Each of these indices is computed by performing the ${\rm argmax(\cdot)}$ operation at the output probability vector of each fourth MBACNN module at each $k$-th RIS agent, and essentially represents the preference of this agent for the precoding selection based on its individual, local, channel observations. Note, however, that none of the RIS agents can directly control the TX precoder. Instead, the $(K+1)$-th agent receives as input the $K$-dimensional vector $\mathbf{\bar{v}}(t) \triangleq [\bar{v}_1(t)\,\bar{v}_2(t)\,\cdots\,\bar{v}_K(t)]$, including the precoding vector indices selected by all RIS agents, and outputs the final precoder to be selected. This operation is handled by the very small FF NN $g_\mathbf{w}^{(5)}(\cdot)$, whose output is a $|\mathcal{V}|$-dimensional vector of probabilities so that $\mathbf{v}(t)$ is sampled/decided accordingly; this is implemented with a module same as the fourth module of the MBACNN. The motivation behind this NN is for the pattern recognition operations to be handled locally at each RIS agent who has more extensive CSI, allowing the $(K+1)$-th agent at the RX to learn simplified strategies. For example, $g_\mathbf{w}^{(5)}(\cdot)$ may converge to a typical majority voting function, or may learn to preferentially select the decisions of certain RISs due to environmental factors (e.g., TX or RX proximity).

During the deployment of the proposed online learning approach, each of the $K$ RIS agents hosts the initial four modules of the MBACNN, and the $(K+1)$-th agent at the RX side, tasked with TX precoder selection, only implements the final module's MLP. The entire architecture is evolved via CoSyNE as a single NN, with the fitness function $f$ calculated as described in Algorithm~\ref{alg:fitnessDist}. In contrast to the centralized setup discussed previously in Section~\ref{cen_vs_dis}, only two low-dimensional vectors are required to be exchanged during operation time: \textit{i}) $\bar{v}_k(t)$ from each $k$-th RIS agent to the RX agent, as well as \textit{ii}) $\mathbf{h}(t)$ from either the TX or the RX (depending on the node selected to receive the pilots during channel estimation), which may be broadcasted toward all RISs simultaneously. Admittedly, this framework further involves broadcasting a complete set of the NN parameter vectors $\mathbf{w}$ from the computational hardware node to all $K$ RIS agents before every fitness function evaluation step of the training algorithm. Nevertheless, this overhead is only incurred during training. In a deployment scenario, the NN weights would remain fixed, therefore, no such messages are needed.

\SetKwInput{KwData}{Input}
\begin{algorithm}[!t]
\caption{Fitness Function for Multiple RISs}\label{alg:fitnessDist}
\KwData{Horizon $T \geq 0$, candidate MBACNN $g^{(1:4)}_\mathbf{w}(\cdot)$, candidate TX NN $g^{(5)}_\mathbf{w}(\cdot)$,  noise level $\sigma^2$, and TX power level $P$.}
Set $f \gets 0$. \\
\For{$t_e=1,2,\ldots,T_E$}{
\For{$t=1,2,\ldots,T$}{
Estimations for $\mathbf{h}(t)$ and $\mathbf{H}_{1,k},(t)$, $\mathbf{h}_{2,k}(t)$ $\forall k$.\\
\For{$k=1,2,\ldots,K$}{
Obtain $\{ \boldsymbol{\phi}_k(t), \bar{v}_k(t) \} = g_\mathbf{w}^{(1:4)}\left(\mathbf{h}(t),\mathbf{H}_{1,k}(t),\mathbf{h}_{2,k}(t)\right)$.\\
}
Construct $\mathbf{\bar{v}}(t) = [\bar{v}_1(t)
\,\bar{v}_2(t)\,\cdots\,\bar{v}_K(t)]$. \\
Set $\mathbf{v}(t)= g_\mathbf{w}^{(5)}(\mathbf{\bar{v}}(t))$. \\
Set $f \gets f+\gamma\left(g_{\mathbf{w}}\left(\mathbf{h}(t), \{\mathbf{H}_{1,k}(t)\}_{k=1}^K,\{\mathbf{h}_{2,k}(t)\}_{k=1}^K\right)\right)$.
}}
Set $f \gets \frac{f}{T_ET}$.\\
Return $f$.
\end{algorithm}

 \section{Numerical Results and Discussion}
In this section, we present performance evaluation results for the proposed MBACNN framework, with parameters as described in Section~\ref{sec:arch} and the corresponding NE-based training procedure, for the online configuration of RIS-empowered MISO communication systems. We have considered both stochastic and geometrical channel models, as well as smart environments with a single and multiple RISs.
The user-defined parameters of the CoSyNE algorithm have been empirically set as listed in Table~\ref{tab:cosyneParams}.

\subsection{Benchmarks}
Three benchmarks were simulated for the single-RIS (i.e., single-agent) case: the DRL agent as~\cite{DRL_RIS_secure,RIS_MISO_exploiting}, a feed forward (FF) NE-based agent, and a classical discrete optimization scheme ~\cite{introtoEA}. Since recent work on similar problems proposes policy gradient heuristics with FF NNs~\cite{Stylianop1Bit,AlexandroPervasive,DRL_RIS_secure,RIS_MISO_exploiting}, we have considered an Advantage Actor Critic (A2C) algorithm~\cite{asynchronousForDRL}. An implementation with $4$ hidden layers of $400$ units has been used, which was trained for $15000$ episodes for every trial. We have also experimented with two other variants, namely, proximal policy optimization~\cite{schulmanPPO} and simple deep actor critic, without notable improvement, hence, we decided to exclude from our performance demonstrations.

To investigate whether the performance improvement is attributed to the proposed NE-based training algorithm or to the proposed NN architecture, a simple FF NN was trained with the same evolutionary algorithm as our MBACNN.
That NN comprised of five hidden layers with $800$, $600$, $600$, $500$, and $200$ neurons, respectively, containing approximately $1.2 \times 10^8$ trainable parameters for a typical system setting in the results that follow. It is noted that, for the same setting of parameters, the proposed MBACNN contained fewer than $9.2 \times 10^5$ weights overall.

The discrete optimization baseline used, termed as ``Lightweight Genetic Algorithm'' (LGA) in the following figures, works as follows. At each time instant $t$, we ran, for each possible TX precoding vector, a lightweight genetic algorithm with $15$ individuals representing candidate RIS phase configurations for $5$ generations, and kept the best precoding and RIS configuration pair. The default genetic algorithm of pygad~\cite{pygad} was used. This methodology has an increased computational complexity since it searches among all precoding combinations and runs the genetic algorithm to completion at every time step, in contrast to the parallelizable operations involved by the NN-based approaches, therefore, its application is expected to be limited in practice.

\begin{table}[!t]
    \centering
    \caption{Parameters used for the CoSyNE algorithm.}
    \begin{tabular}{|c|c|}
    \hline
    \hline
    Parameter & Value\\
    \hline\hline
      Population size $L_{\rm pop}$   & 100 \\ \hline
      Mutation standard deviation $\sigma_{\rm mut}$ &0.2 \\ \hline
      Mutation probability $p_{\rm mut}$ & 0.3 \\ \hline
      Number of generations& 25\\ 
      \hline
      \hline
    \end{tabular}
    \label{tab:cosyneParams}
\end{table}
\begin{figure*}
    \centering
    \begin{subfigure}{.3\textwidth}
        \centering
        \includegraphics[width=\textwidth]{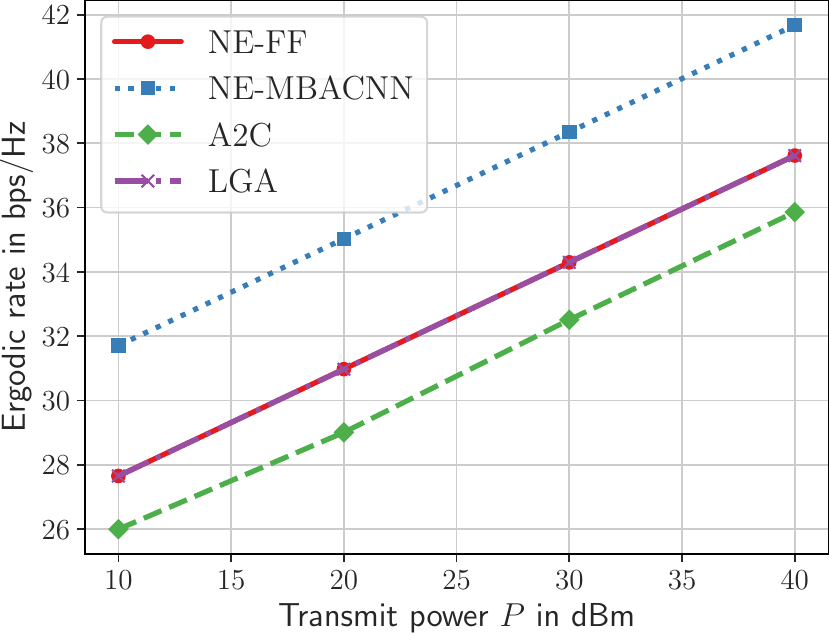}
        \caption{Vs. TX power levels.}\label{subfig:snr_duel}
    \end{subfigure} \hspace{0.25cm}
    \begin{subfigure}{.3\textwidth}
        \centering
        \includegraphics[width=\textwidth]{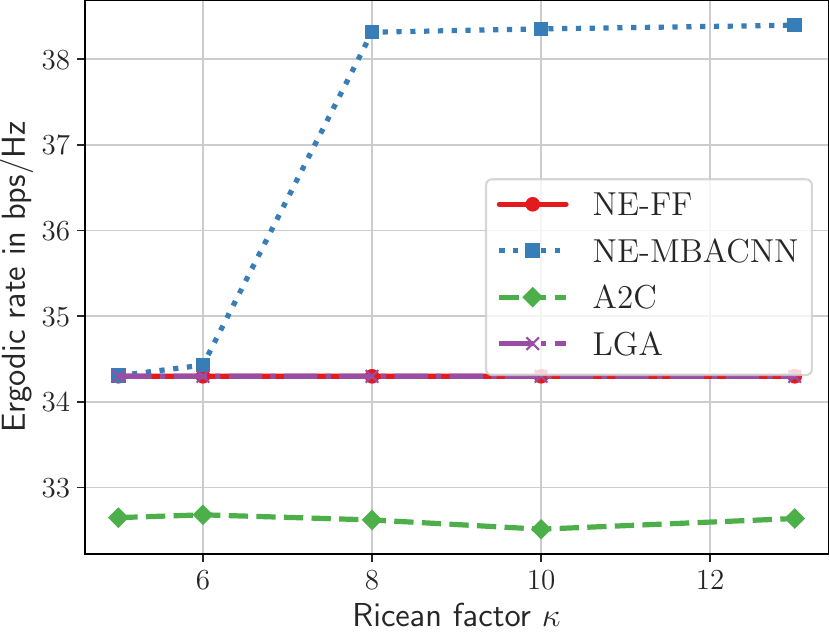}
        \caption{Vs. Ricean factors of the TX-RIS link.}\label{subfig:snr_k}
    \end{subfigure} \hspace{0.25cm}
    \begin{subfigure}{.3\textwidth}
        \centering
        \includegraphics[width=\textwidth]{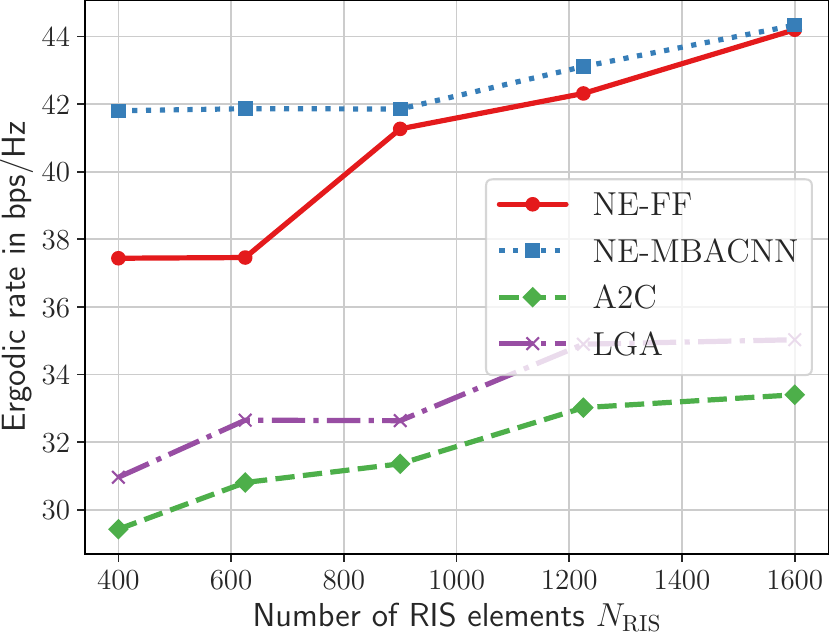}
        \caption{Vs. numbers of RIS elements.}\label{subfig:snr_elem}
    \end{subfigure}%
    \caption{Achievable rate with the proposed MBACNN framework versus (vs.) different system and channel parameters.}\label{fig:resStat1}
\end{figure*}

For the multi-RIS (i.e., multi-agent) case, we considered two benchmarks. First, we evolved an NE-FF with similar structure to the respective single-RIS benchmark, which was enriched with a linear layer of $16$ hidden units for the final TX precoding selection, as described in Section~\ref{sec:distr}. This was optimized to maximize the fitness function described in Algorithm~\ref{alg:fitnessDist}. Second, we evolved a CENTralized NE-FF (NE-FF-CENT) NN having the same number of hidden layers and units as the respective single-RIS benchmark. This variation received as input all channel matrices and gave as output the phase configurations for all RISs, while also selecting the TX precoding vector.

\subsection{Results with Stochastic Channels}
We have considered the TX positioned at the Cartesian coordinates $(0, 0, 2.0)$~m, the RIS placed in Line Of Sight (LOS) with the TX at the position $(0,3,2.0)$~m, and the RX at the position $(8,10,1.5)$~m. The number of TX antennas was set as $N_{\rm TX}=16$ and the noise level to $-50$ dBm. The channel between the RX and the RIS, i.e., $\mathbf{h}_2(t)$, was modeled as a Ricean fading channel with parameter $\kappa$, while the direct TX-RX channel was assumed as totally blocked due to the presence of obstacles. The RIS was modeled as a uniform planar array of equal number of elements in each row and column, while the TX antennas formed a horizontal uniform linear array. The precoding vector codebook $\mathcal{V}$ at the TX consisted of the columns of the $N_{\rm{TX}} \times N_{\rm{TX}}$ DFT matrix, and the time horizon was fixed to $T=50$.

In Fig.~\ref{fig:resStat1}, we have evaluated the achievable ergodic rate for all considered system design approaches as $\log_2\left(1+\gamma(t)\right)$, 
averaging over $T_E=20$ additional episodes  (i.e., a total of $20\times 50=1000$ random channels) after the completion of each training procedure. It can be inferred from Fig.~\ref{subfig:snr_duel}, where we have set $\kappa=10$~dB, $N_{\rm RIS}=400$, and varied the TX power budget $P$, that the proposed NE-optimized MBACNN (NE-MBACNN) greatly outperforms all benchmarks under all $P$ values. Aiming to investigate the generalization of our proposed approach across different channel conditions in Fig.~\ref{subfig:snr_k}, we have set $P$ to $30$~dBm and varied the Ricean $\kappa$-factor. As observed, NE-MBACNN outperforms the A2C and LGA benchmarks for all considered $\kappa$ values, while it achieves similar performance to the simple FF NN for up to $\kappa=6$~dB, i.e., for relatively rich scattering conditions. Yet, the proposed scheme performs substantially better that the simple FF for larger $\kappa$ values, where the channel conditions  becomes LOS-dominant, and consequently, the RIS has the potential for effective beam steering policies\cite{NearFieldBeamManag}. This behavior is attributed to the fact that, in rich scattering, the correlations among the elements of the channel matrices, such as those illustrated in Fig.~\ref{fig:riceanSeqDemo}, become less discernible, or cease to exist completely, due to the increased contribution of the stochastic non-LOS component of the Ricean channel model.

Finally, in Fig.~\ref{subfig:snr_elem}, we have investigated the impact of the RIS size on the rate performance by setting $P=20$ dBm and $\kappa=10$. It is shown that, for $N_{\rm RIS}\leq1000$, NE-MBACNN immensely outperforms all baseline schemes. It can be also seen that NE-MBACNN's gap with NE-FF is reduced from $N_{\rm RIS}=1000$ and on. We attribute this behavior to the two orders of magnitude more parameters contained in the NE-FF scheme compared to our NE-MBACNN approach.
Under a deployment scenario of also large number $N_{\rm TX}=1000$ of TX antenna elements, the number of trainable parameters in the final NN should scale accordingly, however, we have kept the NN architecture fixed to facilitate comparisons.


\subsection{Results with Geometrical Channels}
The performance of the proposed NE-MBACNN scheme, in comparison with the considered benchmarks, was also studied under a geometric channel model. In particular, we have utilized the popular ``DeepMIMO'' dataset~\cite{deepMimo}, sampling channel realizations from the outdoor ``O1'' scenario. In this setting, ``BS 3'' was used as the TX, ``BS 8'' played the role of the RIS, and different channel realizations were emulated by sampling random user positions inside ``User Grid 2.'' In that way, sampling over different user positions was seen as averaging over a range of different positions, antenna/RIS orientations, and pathloss with correlated characteristics. The number of paths was fixed to $15$ and a narrowband carrier at $28$~GHz was used. We have assumed that, at each time instant $t$, the TX wishes to communicate with a user uniformly sampled from the latter grid, and the RIS needs to modify the discrete phase configuration of its elements to facilitate data communication.

\begin{figure}
    \centering
    \begin{subfigure}{0.35\textwidth}
         \scalebox{0.4}{\includegraphics{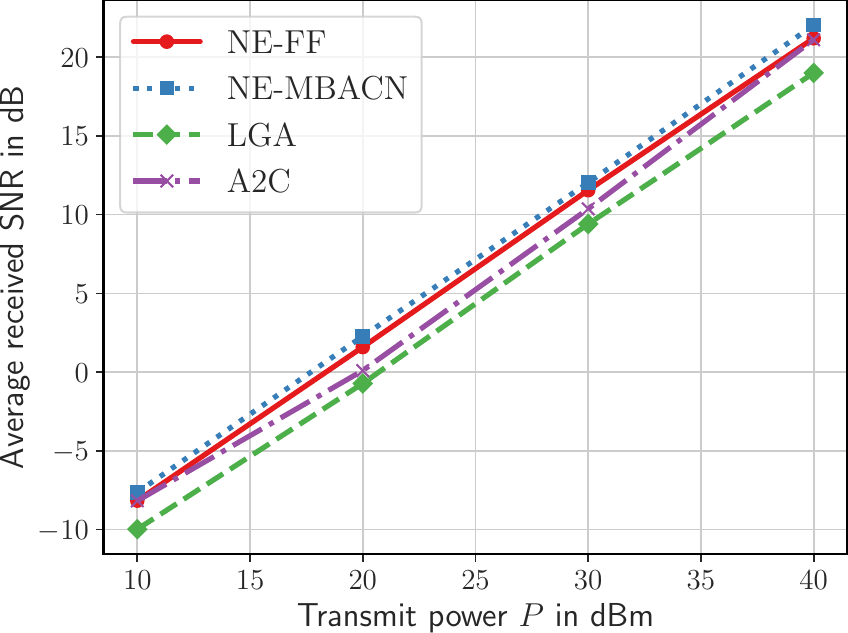}}\caption{$N_{\rm TX}=16$.}
    \end{subfigure}\hspace{2.25cm}
\begin{subfigure}{.35\textwidth}
         \scalebox{0.4}{\includegraphics{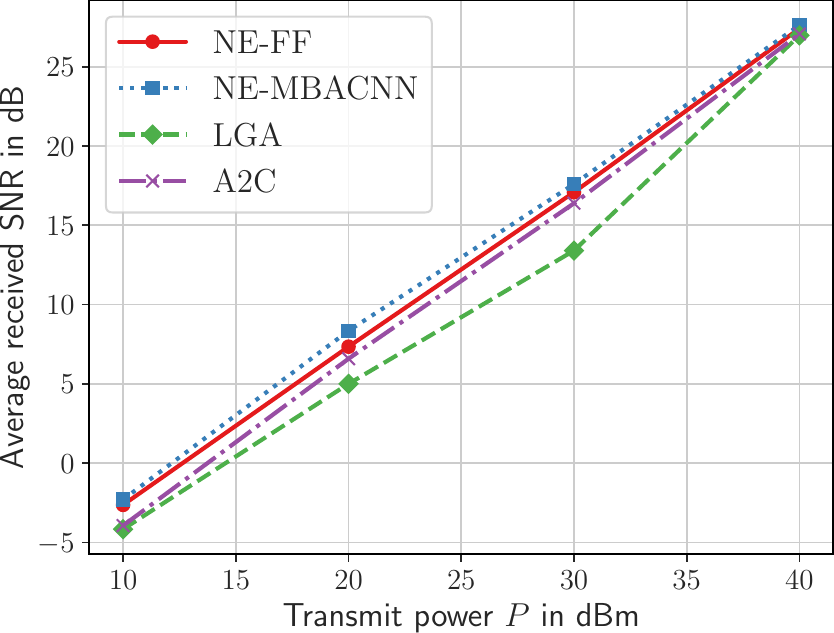}}\caption{$N_{\rm TX}=32$.}
    \end{subfigure}
    \caption{Received SNR with the proposed MBACNN framework vs. TX power level for the considered geometric channels.}
    \label{fig:resGEOM}
\end{figure}
\begin{figure*}
    \centering
    \begin{subfigure}{.3\textwidth}
        \centering
        \includegraphics[width=\textwidth]{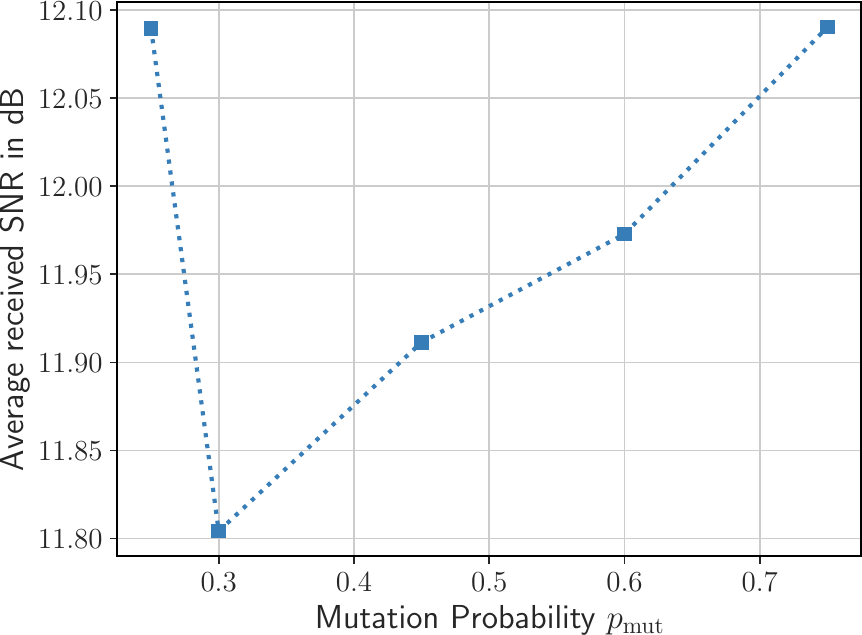}
        \caption{Vs. $p_{\rm mut}$ values.}
    \end{subfigure}
    \hspace{0.25cm}
    \begin{subfigure}{.3\textwidth}
        \centering
        \includegraphics[width=\textwidth]{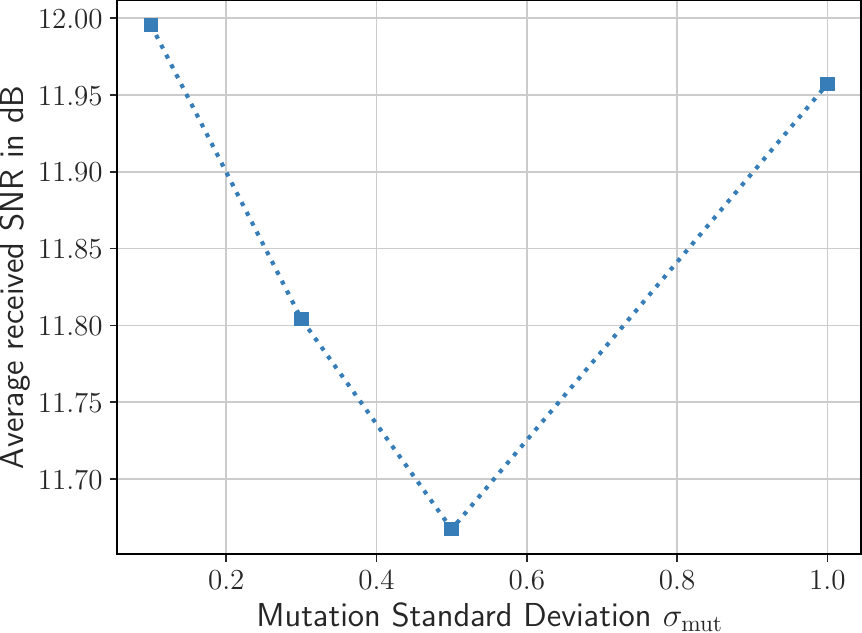}
        \caption{Vs. $\sigma_{\rm mut}$ values.}
    \end{subfigure}
    \hspace{0.25cm}
    \begin{subfigure}{.3\textwidth}
        \centering
        \includegraphics[width=\textwidth]{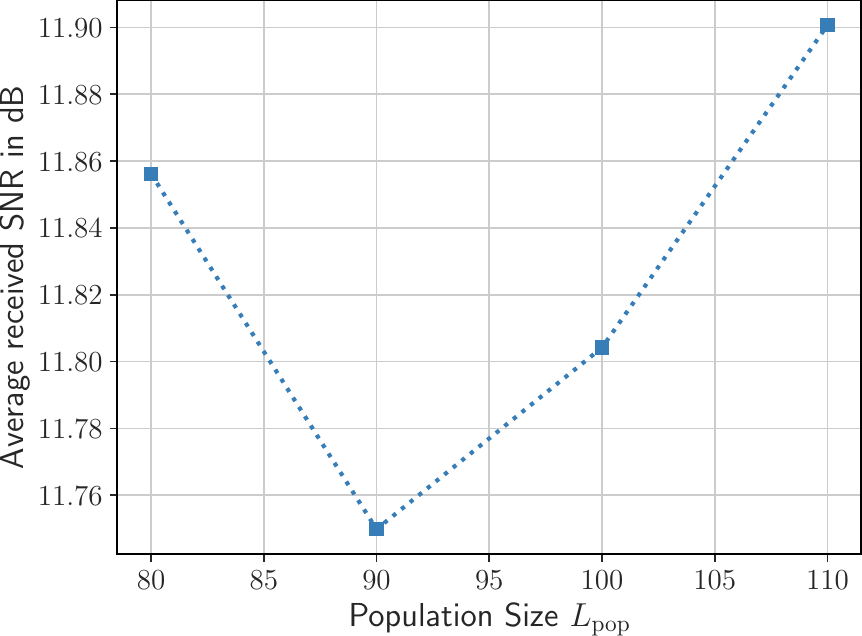}
        \caption{Vs. $L_{\rm pop}$ values.}
    \end{subfigure}%
    \caption{Sensitivity analysis of the proposed NE-MBACNN scheme for the considered geometric channels.}\label{fig:sensitivity}
\end{figure*}

Considering $N_{\rm TX}$ equal to $16$ and $32$, and $T_E=20$ episodes, we fixed the noise level $\tilde{n}(t)$ to $-50$~dBm and varied $P$ in Fig.~\ref{fig:resGEOM} to illustrate the average received SNR with all studied system design schemes. It is evident that the proposed NE-MBACNN scheme can be evolved to obtain better designs/policies than all benchmarks. To verify the robustness of our scheme, we also conducted an extensive sensitivity analysis on this geometric channel set up. For $P=30$ dBm, we first varied the mutation probability $p_{\rm mut}$ from $0.25$ to $0.75$, then, the mutation standard deviation $\sigma_{\rm mut}$ from $0.1$ to $1$ (see \eqref{eq:mutation}), and finally the population size $L_{\rm pop}$ from $80$ to $110$. At each test, all other parameters were kept fixed. The average SNR for each study is given in Fig.~\ref{fig:sensitivity}, where it becomes apparent that NE-MBACNN is robust to small changes in the hyper-parameters; there is at most $10\%$ difference in the scheme's objective value for different parameters. 

\begin{figure}
\centering
    \scalebox{0.4}{\includegraphics[]{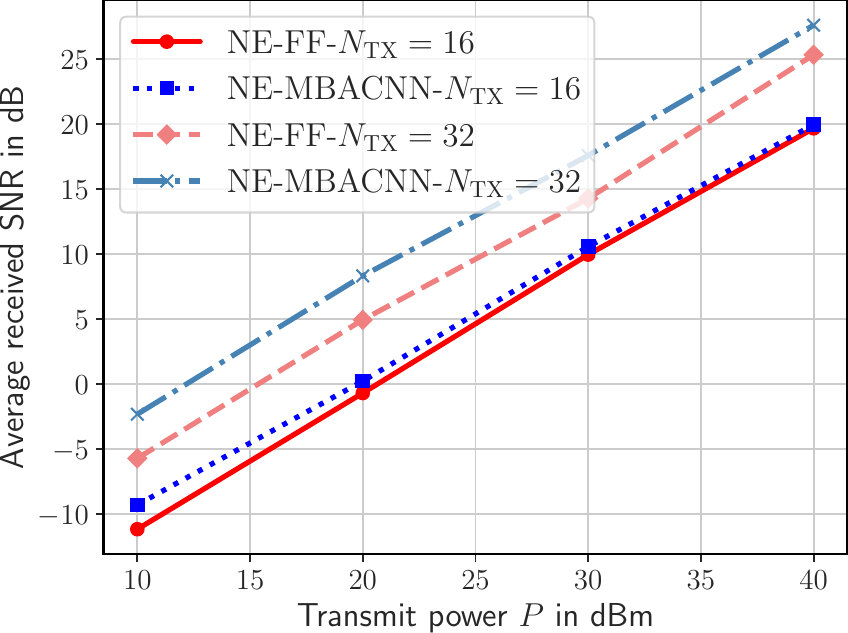}}
    \caption{Generalization of the proposed NE-MBACNN scheme considering the model in~\eqref{eq:gen} with $\epsilon=0.1$ and $\alpha= 1/3$.}
    \label{fig:geomGen}
\end{figure}

Finally, to examine how competent  our NE-MBACNN scheme is on generalizing to slightly different testing environments, we performed the following investigation. We generated a slightly different wireless environment by adding AWGN to each RIS-RX channel vector $\mathbf{h}_2(t)$, in particular, we replaced this vector in \eqref{eq:setUP} and \eqref{eq:SNR-DEF} with the following one: 
\begin{equation}\label{eq:gen}
    \hat{\mathbf{h}}_{2}(t)= \mathbf{h}_{2}(t)+\epsilon^2 \mathcal{C N}\left(\mathbf{0}_{N_{\rm RIS}},\sigma^2_n \mathbf{I}_{N_{\rm RIS}}\right),
\end{equation}
where the noise variance is a multitude of the channel magnitude, i.e., $\sigma^2_n=\alpha|\mathbf{h}_{2}(t)|$. We have set $\epsilon=0.1$ and $\alpha=1/3$ in Fig.~\ref{fig:geomGen} and repeated the experiment with the geometric setup of Fig.~\ref{fig:resGEOM}. As shown, our NE-MBACNN can lead to higher received SNR values than the NE-optimized FF scheme, with this difference being especially prevalent for low $P$ values. Our deployed attention and convolution modules can extract important features regarding the underlying channel structures, therefore, RIS phase configuration policies with better generalization capabilities are achievable. It is noted that this AWGN-based investigation was designed to roughly capture the mismatch of channel parameters between the training and evaluation environments, and may additionally be thought to be akin to channel-estimation-induced errors.
Even though performing a complete generalization study accounting for all sources of variability and errors in a realistic wireless system falls beyond the scope of this paper, we believe that this synthetic setup is adequate in showcasing that our NE-MBACNN scheme exhibits better generalization properties compared to the considered NE-FF benchmark.

\subsection{Results for Multi-RIS Deployments}
\begin{figure}
    \centering \scalebox{0.4}{
\includegraphics[]{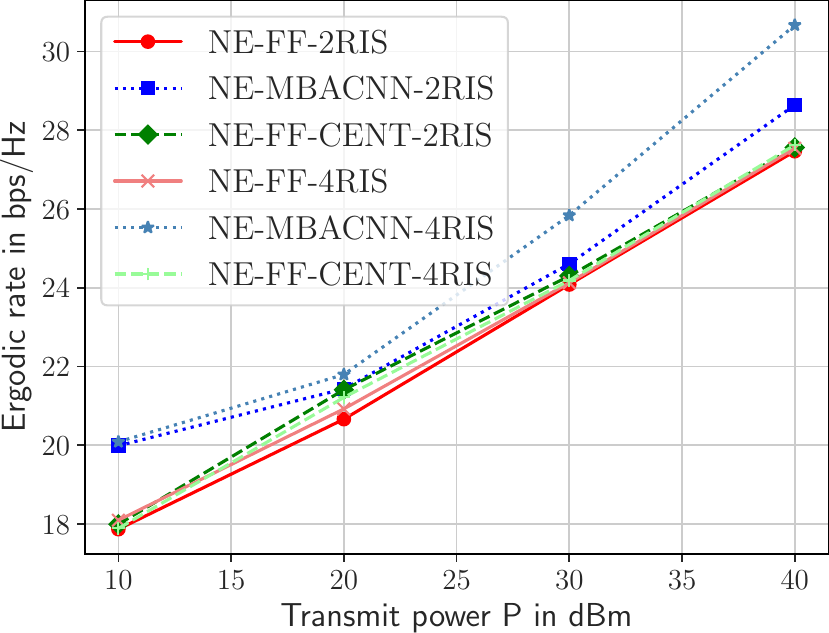}}
    \caption{Achievable rate with the proposed NE-MBACNN scheme for $K=2$ and $4$ RISs vs. TX power levels.}
    \label{fig:multi_ris_res}
\end{figure}

We now consider setups with $K=2$ and $4$ RISs, each having $N_{\rm RIS}=400$ elements. In the first setup, two RISs are positioned at the points $(3, 3, 2)$~m and $(6,6,-2)$~m. For the second setup, two additional RISs are positioned at the points $(3, 3, -2)$~m and $(6,6,2)$~m. The TX equipped with $N_{\rm TX}=16$ antennas was positioned at $(0, 0, 2.0)$~m and the single-antenna RX at $(10,10,5)$~m. We assumed that the direct TX-RX channel is Ricean faded with $\kappa=10$~dB and an attenuation of $10$~dBm. All TX-RIS channels were LOS and all RIS-RX channels were Ricean ones with $\kappa=10$~dB, while the noise variance was set to $-50$~dBm. The achievable rate vs. the TX power, which has been varied from $10$ to $40$~dBm, is illustrated in Fig.~\ref{fig:multi_ris_res}. The structure of the proposed MBACNN for both multi-RIS cases was the same with the previous experiments, enriched with an FF NN with a single hidden layer of $16$ units for the fifth module, as discussed in Section~\ref{sec:distr}. It can be observed from Fig.~\ref{fig:multi_ris_res} that our distributed NE-MBACNN scheme achieves larger SNR values than both NE-FF and the NE-FF-CENT despite having significantly fewer trainable parameters. As shown, this performance superiority is more pronounced for the case of $K=4$ RISs. 
 
\section{Conclusions}
In this paper, we presented a novel MBACNN architecture for the online joint
configuration of the RIS discrete phase responses and the codebook-based TX precoder in RIS-empowered MISO communication systems. 
The proposed architecture consists of separate attention branches followed by a deep convolutional module and two deep FF modules. The optimization of the MBACNN parameters was performed with NE, which can overcome the problem of non-differentiability, introduced by the RIS discrete responses. This methodology was then extended to multi-RIS-empowered MISO communication systems where a distributed and cooperative training and operating procedure was designed, in the effort of reducing the information exchange overheads. Our performance evaluation results showcased that the proposed design framework outperforms existing DRL schemes, discrete optimization benchmarks, and FF NNs with vastly larger numbers of trainable parameters. In addition, the included ablation and generalization studies demonstrated that the proposed NE-based training procedure is rather robust to changes in its hyper-parameters and the testing wireless environment.


\bibliographystyle{ieeetr}
\bibliography{references}

\end{document}